%% file: main.tex
\documentclass{article}
\input{math_commands}

\usepackage{microtype}
\usepackage{graphicx}
\usepackage{subfigure}
\usepackage{booktabs} 
\usepackage{url}
\usepackage{booktabs}
\usepackage{siunitx}
\usepackage{wrapfig}

\newcommand{\exampleswebpage}{\url{https://goo.gl/
magenta/music-transformer-autoencoder-examples}}
\newcommand{\dataset}{YouTube }

\usepackage{hyperref}

\usepackage[accepted]{icml2020}

\icmltitlerunning{Encoding Musical Style with Transformer Autoencoders}

\begin{document}

\twocolumn[
\icmltitle{Encoding Musical Style with Transformer Autoencoders}



\icmlsetsymbol{thanks}{*}

\begin{icmlauthorlist}
\icmlauthor{Kristy Choi}{stan,thanks}
\icmlauthor{Curtis Hawthorne}{goo}
\icmlauthor{Ian Simon}{goo}
\icmlauthor{Monica Dinculescu}{goo}
\icmlauthor{Jesse Engel}{goo}
\end{icmlauthorlist}

\icmlaffiliation{stan}{Department of Computer Science, Stanford University *Work completed during an internship at Google Brain}
\icmlaffiliation{goo}{Google Brain}

\icmlcorrespondingauthor{Kristy Choi}{kechoi@cs.stanford.edu}

\icmlkeywords{Machine Learning, ICML}

\vskip 0.3in
]



\printAffiliationsAndNotice{}  

\begin{abstract}
\input{sections/abstract}
\end{abstract}

\input{sections/introduction}
\input{sections/prelims}
\input{sections/conditional}
\input{sections/metric}
\input{sections/experiments}
\input{sections/related}
\input{sections/conclusion}
\input{sections/acks}

\bibliography{references}
\bibliographystyle{icml2020}
\raggedbottom

\pagebreak
\input{sections/supplement}

\end{document}

%% file: math_commands.tex
\usepackage{amsmath,amsfonts,bm}

\def\Figref#1{Figure~\ref{#1}}

\def\eqref#1{equation~\ref{#1}}

\def\1{\bm{1}}

\DeclareMathAlphabet{\mathsfit}{\encodingdefault}{\sfdefault}{m}{sl}
\SetMathAlphabet{\mathsfit}{bold}{\encodingdefault}{\sfdefault}{bx}{n}



%% file: sections/abstract.tex
We consider the problem of learning high-level controls over the global structure of generated sequences, particularly in the context of symbolic music generation with complex language models. In this work, we present the Transformer autoencoder, which aggregates encodings of the input data across time to obtain a global representation of style from a given performance. We show it is possible to combine this global representation with other temporally distributed embeddings, enabling improved control over the separate aspects of performance style and melody. Empirically, we demonstrate the effectiveness of our method on various music generation tasks on the MAESTRO dataset and a YouTube dataset with 10,000+ hours of piano performances, where we achieve improvements in terms of log-likelihood and mean listening scores as compared to baselines.

%% file: sections/introduction.tex
\section{Introduction}
\label{intro}
\begin{figure*}
\centering
\includegraphics[width=.8\textwidth]{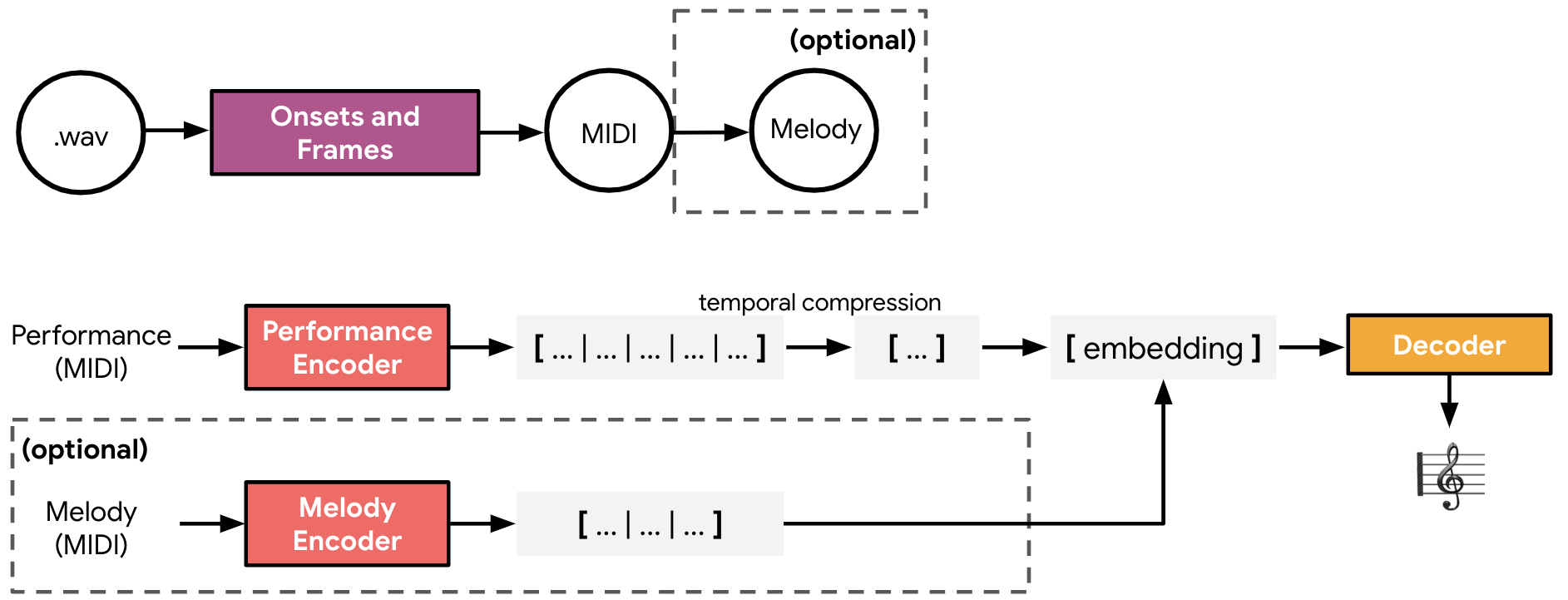}
\caption{A flowchart of the Transformer autoencoder. We first transcribe the \texttt{.wav} data files into MIDI using the Onsets and Frames framework, then encode them into performance representations to use as input. The output of the performance encoder is then aggregated across time and (optionally) combined with a melody embedding to produce a representation of the entire performance, which is then used by the Transformer decoder at inference time.}
\label{fig:model_flowchart}
\end{figure*}

There has been significant progress in generative modeling, particularly with respect to creative applications such as art and music \citep{oord2016wavenet,engel2017neural,ha2017neural,huang2019counterpoint,payne2019}. As the number of generative applications increase, it becomes increasingly important to consider how users can interact with such systems, particularly when the generative model functions as a tool in their creative process \citep{engel2017latent,gillick2019learning} To this end, we consider how one can learn high-level controls over the global structure of a generated sample. We focus on the domain of symbolic music generation, where Music Transformer \citep{huang2018music} is the current state-of-the-art in generating high-quality samples that span over a minute in length.

The challenge in controllable sequence generation is twofold. First, Transformers \citep{vaswani2017attention} and their variants excel as unconditional language models or in sequence-to-sequence tasks such as translation, but it is less clear as to how they can: (1) \textit{learn} and (2) \textit{incorporate} global conditioning information at inference time. This contrasts with traditional generative models for images such as the variational autoencoder (VAE) \citep{kingma2013auto} or generative adversarial network (GAN) \citep{goodfellow2014generative}, both of which can incorporate global conditioning information (e.g. one-hot encodings of class labels) as part of their training procedure \citep{sohn2015learning,sonderby2016ladder,isola2017image,van2016conditional}. Second, obtaining a "ground truth" annotation that captures all the salient features of a musical performance may be a prohibitively difficult or expensive task that requires domain expertise  \cite{bertin2011automatic}. Thus even if conditioning was straightforward, the set of performance features that will be relevant for synthesizing a desired sample will remain ambiguous without descriptive tags to guide generation.

In this work, we introduce the Transformer autoencoder, where we aggregate encodings across time to obtain a holistic representation of the performance style in an unsupervised fashion. We demonstrate that this learned global representation can be incorporated with other forms of structural conditioning in two ways. First, we show that given a performance, our model can generate samples that are similar in style to the provided input. Then, we explore different methods to combine melody and performance representations to harmonize a new melody in the style of the given performance. We validate this notion of "perceptual similarity" through quantitative analyses based on note-based features of performances as well as qualitative user listening studies and interpolations. In both cases, we show that combining both global and fine-scale encodings of the musical performance allows us to gain better control of generation, separately manipulating both the style and melody of the resulting sample without the need for explicit labeling. 

Empirically, we evaluate our model on two datasets: the publicly-available MAESTRO \citep{hawthorne2018enabling} dataset, and a YouTube dataset of piano performances transcribed from 10,000+ hours of audio \citep{simon2019}. 
We find that the Transformer autoencoder is able to generate not only  performances that sound similar to the input, but also accompaniments of melodies that follow a given style. In particular, we demonstrate that our model is capable of adapting to a particular musical style at test time \textit{even in the case} where we have one single input performance. 

%% file: sections/prelims.tex
\section{Preliminaries}
\label{gen_inst}

\subsection{Data Representation for Music Generation}
The MAESTRO \citep{hawthorne2018enabling} dataset consists of over 1,100 classical piano performances, where each piece is represented as a MIDI file. The \dataset performance dataset is comprised of approximately 400K piano performances (over 10,000 hours) transcribed from audio \citep{simon2019}.
In both cases, we represent music as a sequence of discrete tokens, effectively formulating the generation task as a language modeling problem. The performances are encoded using the vocabulary as described in \citep{oore2018time}, which captures expressive dynamics and timing. This performance encoding vocabulary consists of 128 \texttt{note\_on} events, 128 \texttt{note\_off} events, 100 \texttt{time\_shift} events representing time shifts in 10ms increments from 10ms to 1s, and 32 quantized \texttt{velocity} bins representing the velocity at which the 128 \texttt{note\_on} events were played. We provide additional details of the data representation, encoding mechanism, and melody extraction procedure in the supplementary material.

\subsection{Music Transformer}
We build our Transformer autoencoder from Music Transformer, a state-of-the-art generative model that is capable of generating music with long-term coherence \citep{huang2018music}. While the original Transformer uses self-attention to operate over absolute positional encodings of each token in a given sequence \citep{vaswani2017attention}, Music Transformer replaces this mechanism with \textit{relative} attention \citep{shaw2018self}, which allows the model to keep better track of regularity based on event orderings and periodicity in the performance. \cite{huang2018music} propose a novel algorithm for implementing relative self-attention that is significantly more memory-efficient, enabling the model to generate musical sequences that span over a minute in length. For more details regarding the self-attention mechanism and Transformers, we refer the reader to \citep{vaswani2017attention,parmar2018image}.

%% file: sections/conditional.tex
\section{Generation with Transformer Autoencoder}
\label{c_generation}
\subsection{Model Architecture}
We leverage the standard encoder and decoder stacks of the Transformer as the foundational building block for our model, with minor modifications that we outline below.

\textbf{Transformer Encoder:} For both the performance and melody encoder networks, we use the Transformer's stack of 6 layers which are each comprised of a: (1) multi-head relative attention mechanism; and a (2) position-wise fully-connected feed-forward network. The performance encoder takes as input the event-based performance encoding of an input performance, while the melody encoder learns an encoding of the melody which has been extracted from the input performance. Depending on the music generation task (Section 3.2), the encoder output(s) are fed into the Transformer decoder. Figure 1 describes the way in which the encoder and decoder networks are composed together.

\textbf{Transformer Decoder:} The decoder shares the same structure as the encoder network, but with an additional multi-head attention layer over the encoder outputs. At each step of the generation process, the decoder takes in the output of the encoder, as well as each new token that was generated in the previous timestep. 

\subsection{Conditioning Tasks and Mechanisms}

\textbf{Performance Conditioning and Bottleneck:} We aim to generate samples that sound ``similar" to a conditioning input performance. We incorporate a bottleneck in the output of the Transformer encoder in order to prevent the model from simply memorizing the input \citep{baldi2012autoencoders}. As shown in Figure~\ref{fig:model_flowchart}, we mean-aggregate the performance embedding across the time dimension in order to learn a global representation of style. This mean-performance embedding is then fed into the autoregressive decoder, where the decoder attends to this global representation in order to predict the appropriate target. Although this bottleneck may be undesirable in sequence transduction tasks where the input and output sequences differ (e.g. translation), we find that it works well in our setting where we require the generated samples to be \textit{similar in style} to the input sequence.

\paragraph{Melody \& Performance Conditioning:} Next, we synthesize any given melody in the style of a \textit{different} performance. Although the setup is similar to the melody conditioning problem in \citep{huang2018music}, we note that we also provide a conditioning performance signal, which makes the generation task more challenging. During training, we extract melodies from performances in the training set as outlined in \citep{waite2016}, quantize the melody to a 100ms grid, and encode it as a sequence of tokens that uses a different vocabulary than the performance representation. For more details regarding the exact melody extraction procedure, we refer the reader to the supplement. We then use two distinct Transformer encoders (each with the same architecture) as in Section 3.1 to separately encode the melody and performance inputs. The melody and performance embeddings are combined to use as input to the decoder.

We explore various ways of combining the intermediate representations: (1) \texttt{sum}, where we add the performance and melody embeddings together; (2) \texttt{concatenate}, where we concatenate the two embeddings separated with a \texttt{stop} token; and (3) \texttt{tile}, where we tile the performance embedding across every dimension of time in the melody encoding. In all three cases, we work with the mean-aggregated representation of the input performance. We find that different approaches work better than others on some dataets, a point which we elaborate upon in Section~\ref{experiments}.

\subsection{Model Training}
\paragraph{Input Perturbation:} In order to encourage the encoded performance representations to generalize across various melodies, keys, and tempos, we draw inspiration from the denoising autoencoder \citep{vincent2008extracting} as a means to regularize the model. For every target performance from which we extract the input melody, we provide the model with a \textit{perturbed} version of the input performance as the conditioning signal. We allow this "noisy" performance to vary across two axes of variation: (1) \texttt{pitch}, where we artificially shift the overall pitch either down or up by 6 semitones; and (2) \texttt{time}, where we stretch the timing of the performance by at most $\pm 5\%$. Then for each new data point during training, a single noise injection procedure is randomly sampled from the cross product of all possible combinations of 12 pitch shift values and 4 time stretch values (evaluated in intervals of 2.5\%). At test time, the data points are left unperturbed. In our experiments, we find that this augmentation procedure leads to samples that sound more pleasing \citep{oore2018time}. 

Finally, the model is trained end-to-end with maximum likelihood: for a given sequence $x$ of length $n$, we maximize $\log p_\theta(x) = \sum_{i=1}^n \log p_\theta(x_i|x_{<i})$ with respect to the model parameters $\theta$. We emphasize that training is conducted in an autoencoder-like fashion. Specifically, for performance-only conditioning, the Transformer decoder is tasked with predicting the same performance provided to the encoder. For melody \& performance conditioning, the Transformer autoencoder is trained to predict a new performance using the \textit{combined} melody + performance embedding, where the loss is computed with respect to the input performance.

\begin{table*}[ht]
\centering 
\begin{tabular}{l|l|l}
\toprule
\textbf{Model variation} & MAESTRO & YouTube \\
\midrule
Unconditional model with rel. attention \citep{huang2018music} & 1.840 & 1.49 \\ 
\midrule
Performance autoencoder with rel. attention (ours) & \textbf{1.799} & \textbf{1.384}\\   
\bottomrule
\end{tabular}
\caption{Note-wise test NLL on the MAESTRO and \dataset datasets. We exclude the performance autoencoder baseline (no aggregation) as it memorized the data ($\text{NLL}=0$). Conditional models outperformed their unconditional counterparts.}
\label{table:perf_nll}
\end{table*}

\begin{table*}[ht]
\centering
\begin{tabular}{l|l|l}
\toprule
\textbf{Model variation} & MAESTRO & YouTube \\
\midrule
Melody-only Transformer with rel. attention \citep{huang2018music} & 1.786 & 1.302\\ 
\midrule
Melody \& performance autoencoder with rel. attention, sum (ours) & \textbf{1.706} & 1.275\\   
Melody \& performance autoencoder with rel. attention, concat (ours) & 1.713 & \textbf{1.237}\\   
Melody \& performance autoencoder with rel. attention, tile (ours) & 1.709 & 1.248\\ 
\bottomrule
\end{tabular}
\caption{Note-wise test NLL on the MAESTRO and \dataset datasets with melody conditioning. We note that \texttt{sum} worked best for MAESTRO, while \texttt{concatenate} outperformed all other baselines for the \dataset dataset.}
\label{table:melody_nll}
\end{table*}

%% file: sections/metric.tex
\section{Performance Similarity Evaluation}
\label{metric}

As the development of a proper metric to quantify both the quality and similarity of musical performances remains an open question \citep{engel2019gansynth}, we draw inspiration from \citep{yang2018evaluation,hung2019improving} to capture the style of a given performance based on eight features corresponding to its pitch and rhythm:

\begin{enumerate}
    \item \textit{Note Density (ND)}: The note density refers to the average number of notes per second in a performance: a higher note density often indicates a fast-moving piece, while a lower note density correlates with softer, slower pieces. This feature is a good indicator for rhythm.
    \item \textit{Pitch Range (PR)}: The pitch range denotes the difference between the highest and lowest semitones (MIDI pitches) in a given phrase.
    \item \textit{Mean Pitch (MP) / Variation of Pitch (VP)}: Similar in vein to the pitch range (PR), the average and overall variation of pitch in a musical performance captures whether the piece is played in a higher or lower octave.
    \item \textit{Mean Velocity (MV) / Variation of Velocity (VV)}: The velocity of each note indicates how hard a key is pressed in a musical performance, and serves as a heuristic for overall volume.
    \item \textit{Mean Duration (MD) / Variation of Duration (VD)}: The duration describes for how long each note is pressed in a performance, representing articulation, dynamics, and phrasing.
\end{enumerate}

\subsection{Overlapping Area (OA) Metric}
To best capture the salient features within the periodic structure of a musical performance, we used a sliding window of 2s to construct histograms of the desired feature within each window. We found that representing each performance with such relative measurements better preserved changing dynamics and stylistic motifs across the entire performance as opposed to a single scalar value (e.g. average note density across the entire performance). 

Similar to \citep{yang2018evaluation,hung2019improving}, we smoothed each feature's histogram by fitting a Gaussian distribution -- this allowed us to learn a compact representation per feature through its mean $\mu$ and variance $\sigma^2$. Then to compare two performances, we computed the Overlapping Area (OA) between the Gaussian pdfs of each feature to quantify their similarity. The OA can be used to pinpoint feature-wise similarities between two performances, while the average OA across all features ($\textrm{OA}_{\textrm{avg}}$) can be used as a scalar-value summary to compare two performances together. We use both variants to quantify how similar two musical performances are in terms of their relative features.

Concretely, suppose we compare two performances A and B for the "pitch range" feature. If we model $A \sim \mathcal{N}(\mu_1, \sigma_1^2)$ and $B \sim \mathcal{N}(\mu_2, \sigma_2^2)$, and let $c$ denote the point of intersection between the two pdfs (assuming without loss of generality that $\mu_1 > \mu_2$), the OA between A and B is:
\begin{equation}
\label{oa_metric}
OA(A,B) = 1 - \textrm{erf}\left(\frac{c-\mu_1}{\sqrt{2}\sigma_1^2}\right) + \textrm{erf}\left(\frac{c-\mu_2}{\sqrt{2}\sigma_2^2}\right)
\end{equation}
where $\textrm{erf}(\cdot)$ denotes the error function $\textrm{erf}(x) = \frac{2}{\sqrt{\pi}}\int_0^x e^{-t^2}dt$. We found that other divergences such as the Kullback-Leibler (KL) divergence and the symmetrized KL were more sensitive to performance-specific features (rather than melody) than the OA.
Empirically, we demonstrate that this metric identifies the relevant characteristics of interest in our generated performances in Section~\ref{experiments}. 

We note that for the melody \& performance conditioning case, we performed similarity evaluations of our samples against the original performance from which the melody was extracted, as opposed to the melody itself. This is because the melody (a monophonic sequence) is represented using a different encoding and vocabulary than the performance (a polyphonic sequence). Specifically, we average two OA terms: (1) OA(source performance of extracted melody, generated sample)  and (2) OA(conditioning performance, generated sample), as our final similarity metric. In this way, we account for the contributions of both the conditioning melody and performance sequence.

%% file: sections/experiments.tex
\begin{table*}[ht]
\centering
\begin{tabular}{l|l|l|l|l|l|l|l|l|l}
\toprule
\textbf{MAESTRO} & ND & PR & MP & VP & MV & VV & MD & VD & Avg\\
\midrule
Performance (ours) & \textbf{0.651} & \textbf{0.696} & \textbf{0.634} & \textbf{0.689} & \textbf{0.693} & \textbf{0.732} & \textbf{0.582} & \textbf{0.692} & \textbf{0.67}\\ 
Unconditional & 0.370 & 0.466 & 0.435 & 0.485 & 0.401 & 0.606 & 0.385 & 0.529 & 0.46 \\
\midrule
\textbf{YouTube Dataset}\\
\midrule
Performance (ours) & \textbf{0.731} & \textbf{0.837} & \textbf{0.784} & \textbf{0.838} & \textbf{0.778} & \textbf{0.835} & \textbf{0.785} & \textbf{0.827} & \textbf{0.80}\\ 
Unconditional & 0.466 & 0.561 & 0.556 & 0.578 & 0.405 & 0.590 & 0.521 & 0.624 & 0.54\\

\bottomrule
\end{tabular}
\caption{Average overlapping area (OA) similarity metrics comparing performance conditioned models with unconditional models. Unconditional and Melody-only baselines are from \citep{huang2018music}. The metrics are described in detail in Section~\ref{metric}. The samples in this quantitative comparison are used for the listener study shown in the left graph of Figure~\ref{fig:listening_tests}.}
\label{table:similiarity_metric_perf}
\end{table*}

\begin{table*}[ht]
\centering
\begin{tabular}{l|l|l|l|l|l|l|l|l|l}
\toprule

\textbf{MAESTRO} & ND & PR & MP & VP & MV & VV & MD & VD & Avg\\
\midrule
Melody \& perf. (ours) & \textbf{0.650} & \textbf{0.696} & 0.634 & 0.689 & \textbf{0.692} & 0.732 & \textbf{0.582} & \textbf{0.692} & \textbf{0.67} \\
Perf-only (ours) & 0.600 & 0.695 & \textbf{0.657} & \textbf{0.721} & 0.664 & \textbf{0.740} & 0.527 & 0.648 & 0.66 \\
Melody-only & 0.609 & 0.693 & 0.640 & 0.693 & 0.582 & 0.711 & 0.569 & 0.636 & 0.64 \\
Unconditional & 0.376 & 0.461 & 0.423 & 0.480 & 0.384 & 0.588 & 0.347 & 0.520 & 0.48\\

\midrule
\textbf{YouTube Dataset} \\
\midrule
Melody \& perf (ours) & \textbf{0.646} & \textbf{0.708} & 0.610 & 0.717 & \textbf{0.590} & \textbf{0.706} & \textbf{0.658} & \textbf{0.743} & \textbf{0.67} \\
Perf-only (ours) & 0.624 & 0.646 & 0.624 & 0.638 & 0.422 & 0.595 & 0.601 & 0.702 & 0.61 \\
Melody-only & 0.575 & 0.707 & \textbf{0.662} & \textbf{0.718} & 0.583 & 0.702 & 0.634 & 0.707 & 0.66 \\
Unconditional & 0.476 & 0.580 & 0.541 & 0.594 & 0.400 & 0.585 & 0.522 & 0.623 & 0.54\\
\bottomrule
\end{tabular}
\caption{Average overlapping area (OA) similarity metrics comparing models with different conditioning. Unconditional and Melody-only baselines are from \citep{huang2018music}. The metrics are described in detail in Section~\ref{metric}. The samples in this quantitative comparison are used for the listener study shown in the right graph of Figure~\ref{fig:listening_tests}.}
\label{table:similiarity_metric_mel}
\end{table*}

\begin{figure*}[ht]
\centering
\subfigure[Relative distance to performance A]{\includegraphics[width=.49\textwidth]{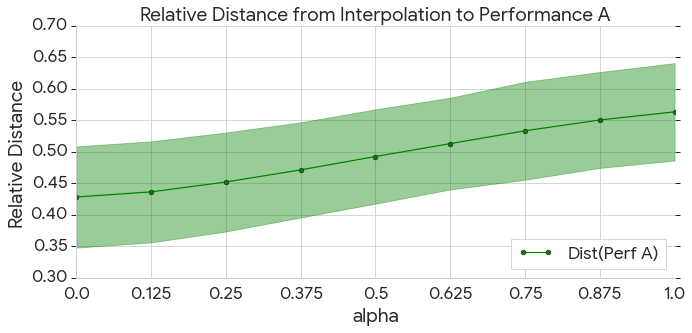}}
\subfigure[Relative distance to melody A]{\includegraphics[width=.49\textwidth]{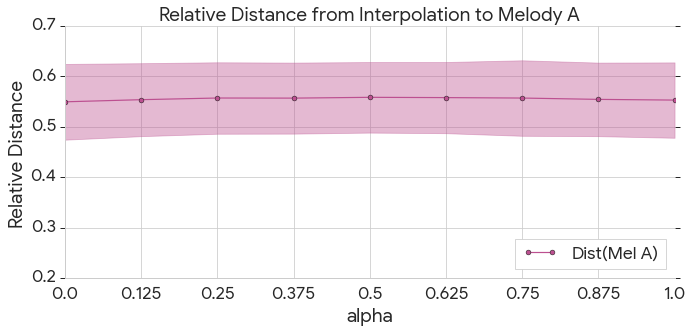}}
\caption{For the \dataset dataset, relative distance from performance A ($\alpha=1$) as $\alpha$ is slowly increased to 1.0 while the conditioned melody is fixed. As in (b), the relative distance to the conditioning melody with respect to a random performance remains fixed while the interpolation is conducted between performances A and B, suggesting that we can control for elements of style and melody separately.}
\label{fig:interp_dist_mel}
\end{figure*}

\section{Experimental Results}
\label{experiments}
\paragraph{Datasets:} We used both the MAESTRO \citep{hawthorne2018enabling} and \dataset datasets \citep{simon2019} for the experimental setup. We used the standard 80/10/10 train/validation/test split from MAESTRO v1.0.0, and augmented the dataset by 10x using pitch shifts of no more than a minor third and time stretches of at most 5\%. We note that this augmentation is distinct from the noise-injection procedure referenced in Section~\ref{c_generation}: the data augmentation merely increases the size of the initial dataset, while the perturbation procedure operates \textit{only} on the input performance signal to regularize the learned model. The \dataset dataset did not require any additional augmentation.

\paragraph{Experimental Setup:} We implemented the model in the Tensor2Tensor framework \citep{vaswani2017attention}, and used the default hyperparameters for training: 0.2 learning rate with 8000 warmup steps, \texttt{rsqrt\_decay}, 0.2 dropout, and early stopping for GPU training. For TPU training, we use AdaFactor with the \texttt{rsqrt\_decay} and learning rate warmup steps to be 10K. We adopt many of the hyperparameter configurations from \citep{huang2018music}, where we reduce the query and key hidden size to half the hidden size, use 8 hidden layers, use 384 hidden units, and set the maximum relative distance to consider to half the training sequence length for relative global attention. We set the maximum sequence length (length of event-based representations) to be 2048 tokens, and a filter size of 1024. We provide additional details on the model architectures and hyperparameter configurations in the supplement.

\subsection{Log-Likelihood Evaluation}
As expected, the Transformer autoencoder with the encoder bottleneck outperformed other baselines. In Tables~\ref{table:perf_nll} and 2, we see that all conditional model variants outperform their unconditional counterparts. For the melody \& performance model, different methods of combining the embeddings work better for different datasets. For example, \texttt{concatenate} led to the lowest NLL for the \dataset dataset, while \texttt{sum} outperformed all other variants for MAESTRO. We report NLL values for both datasets for the perturbed-input model variants in the supplement.

\subsection{Similarity Evaluation}
We use the OA metric from Section~\ref{metric} to evaluate whether using a conditioning signal in both the (a) performance autoencoder (Perf-only) and (b) melody \& performance autoencoder (Melody \& perf) produces samples that are more similar in style to the conditioning inputs from the evaluation set relative to other baselines.

First, we sample 500 examples from the test set as conditioning signals to generate one sample per input. Then, we compare each conditioning signal to: (1) the generated sample and (2) an unconditional sample from the Music Transformer. We compute the similarity metric as in Section~\ref{metric} pairwise and average over 500 examples. As shown in Tables 3 and 4, the performance autoencoder generates samples that have 48\% higher similarity to the conditioning input as compared to the unconditional baseline for the YouTube dataset (45\% higher similarity for MAESTRO).

For the melody \& performance autoencoder, we sample 717*2 distinct performances -- we reserve one set of 717 for conditioning performance styles, and the other set of 717 we use to extract melodies in order to synthesize in the style of a \textit{different} performance. We compare the melody \& performance autoencoder to 3 different baselines: (1) one that is conditioned only on the melody (Melody-only); (2) conditioned only on performance (Perf-only); and (3) an unconditional language model. We find that the Melody \& performance autoencoder performs the best overall across almost all features.

\subsection{Latent Space Interpolations}
Next, we analyze whether the Transformer autoencoder learns a semantically meaningful latent space through a variety of interpolation experiments on both model variants.

\subsubsection{Performance Autoencoder}
We test whether the performance autoencoder can successfully interpolate between different input performances. First, we sample 1000 performances from the \dataset test set (100 for MAESTRO, due to its smaller size), and split this dataset in half. The first half we reserve for the original starting performance, which we call ``performance A", and the other half we reserve for the end performance, denoted as ``performance B." Then we use the performance encoder to embed performance A into $z_A$, and do the same for performance B to obtain $z_B$. For a range $\alpha \in [0, 0.125, \ldots, 0.875, 1.0]$, we sample a new performance $\text{perf}_{\text{new}}$ that results from decoding $\alpha \cdot z_A + (1 - \alpha) \cdot z_B$. We observe how the $\textrm{OA}_{\textrm{avg}}$ (averaged across all features) defined in Section~\ref{metric} changes between this newly interpolated performance $\text{perf}_{\text{new}}$ and performances \{A, B\}.

\begin{figure}[ht]
\centering
\includegraphics[width=.5\textwidth]{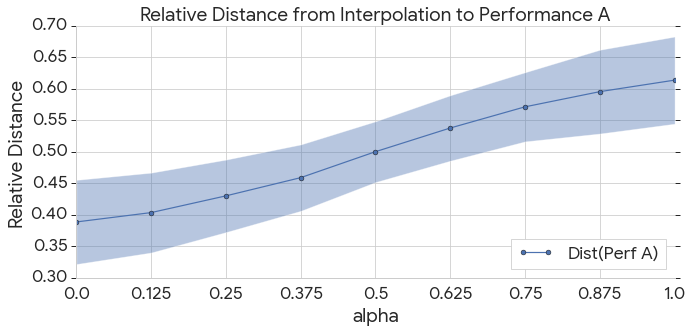}
\caption{For the \dataset dataset, the relative distance from performance A ($\alpha=1$) to the interpolated sample increases as $\alpha$ is slowly increased to 1.0.}
\label{fig:interp_dist}
\end{figure}

Specifically, we compute the similarity metric between each input performance A and interpolated sample $\text{perf}_{\text{new}}$  for all 500 samples, and compute the same pairwise similarity for each performance B. We then compute the normalized distance between each interpolated sample and the corresponding performance A or B, which we denote as: $\texttt{rel\_distance(perf A)} = 1 - \frac{\texttt{OA\_A}}{\texttt{OA\_A + OA\_B}}$, where the OA is averaged across all features. We average this distance across all elements in the set and find in Figure~\ref{fig:interp_dist} that the relative distance between performance A slowly increases as we increase $\alpha$ from 0 to 1, as expected. We note that it is not possible to conduct this interpolation study with non-aggregated baselines, as we cannot interpolate across variable-length embeddings. We find that a similar trend holds for MAESTRO as in Figure~\ref{fig:interp_dist_mel}(a).

\begin{figure*}[ht]
\centering
\subfigure[Performance conditioning study]{\includegraphics[width=.45\textwidth]{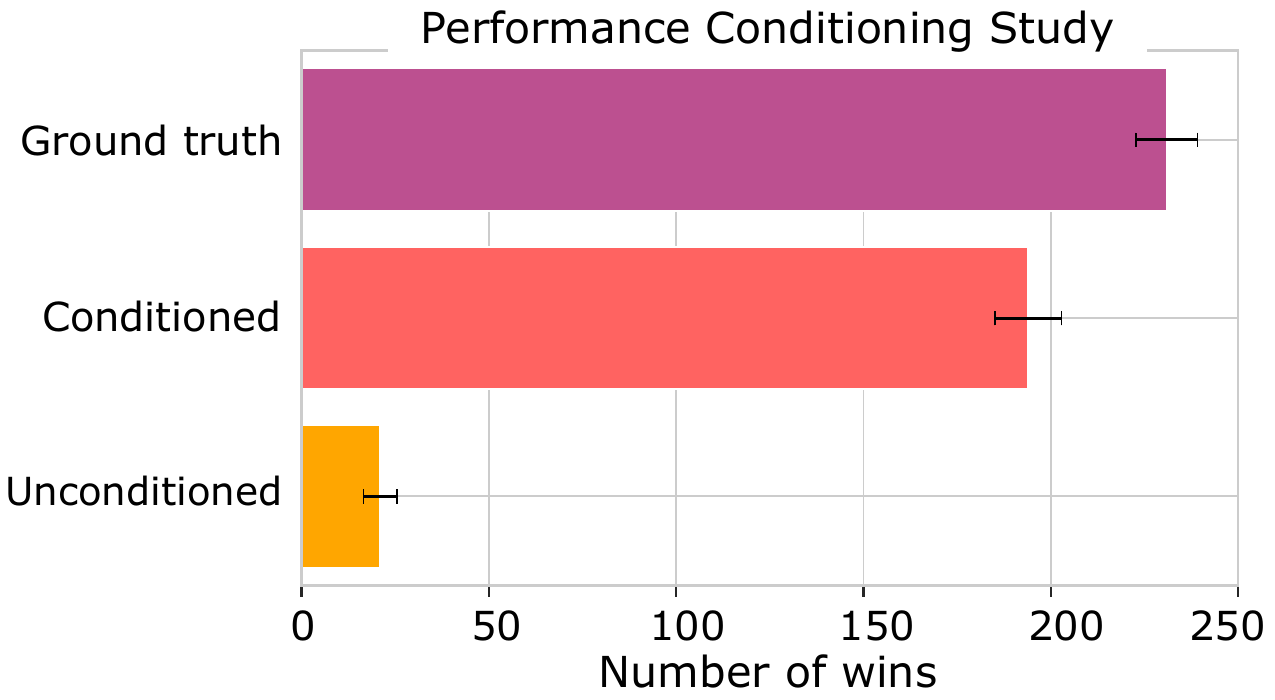}}
\subfigure[Melody conditioning study]{\includegraphics[width=.5\textwidth]{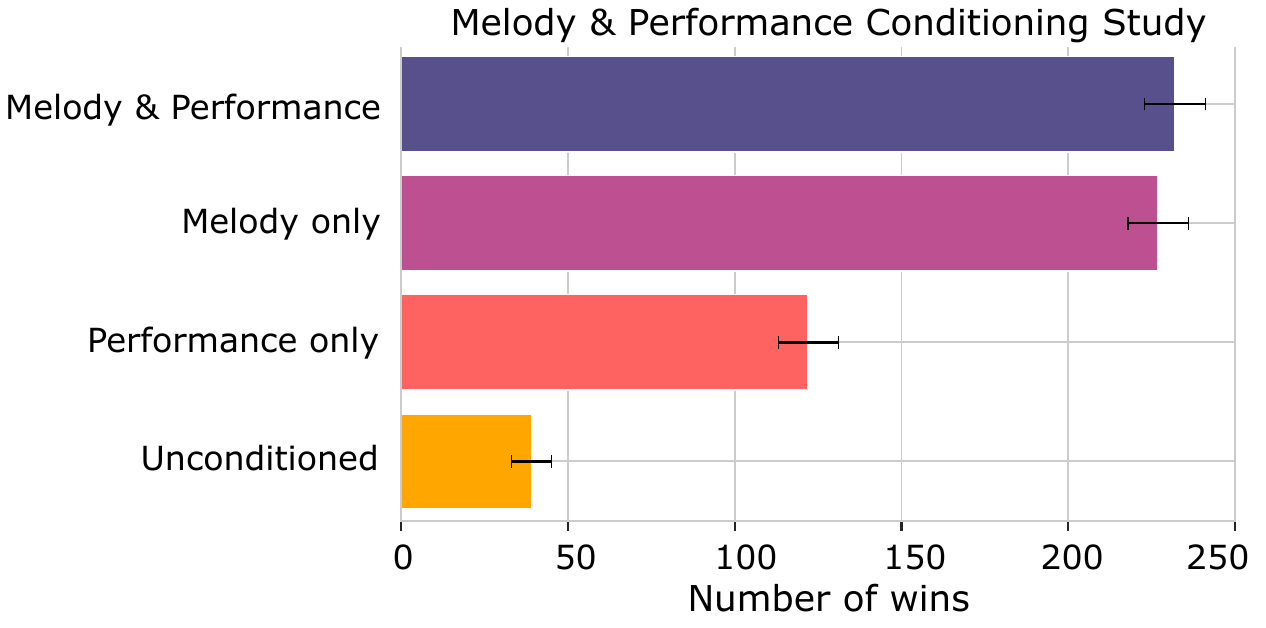}}
\caption{Results of our listening studies, showing the number of times each source won in a pairwise comparison. Black error bars indicate estimated standard deviation of means.}
\label{fig:listening_tests}
\end{figure*}

\subsubsection{Melody \& Performance Autoencoder}
We conduct a similar study as above with the melody \& performance autoencoder. We hold out 716 unique melody-performance pairs (melody is not derived from the same performance) from the \dataset evaluation dataset and 50 examples from MAESTRO. We then interpolate across the different performances, while keeping the conditioning melody input the same across the interpolations.

As shown in Figure~\ref{fig:interp_dist_mel}(a), we find that a similar trend holds as in the performance autoencoder: the newly-interpolated samples show that the relative distance between performance A increases as we increase the corresponding value of $\alpha$. We note that the interpolation effect is slightly lower than that of the previous section, particularly because the interpolated sample is also dependent on the melody that it is conditioned on. Interestingly, in Figure~\ref{fig:interp_dist_mel}(b), we note that the relative distance between the input performance from which we derived the original melody remains fairly constant across the interpolation procedure. This suggests that we are able to factorize out the two sources of variation and that varying the axis of the input performance keeps the variation in melody constant.

\subsection{Listening Tests}
To further evaluate the perceived effect of performance and melody conditioning on the generated output, we also conducted qualitative listening tests. Using models trained on the \dataset dataset, we conducted two studies for separate music generation tasks: one for the performance conditioning, and one for melody and performance conditioning.

\subsubsection{Performance Conditioning}
For performance conditioning, we presented participants with a 20s performance clip from the YouTube evaluation dataset that we used as a conditioning signal. We then asked them to listen to two additional 20s performance clips and to use a Likert scale to rate which one sounded most similar in style to the conditioning signal. The sources rated by the participants included ``Ground Truth" (a different snippet of the same sample used for the conditioning signal), ``Conditioned" (output of the Performance Autoencoder), and ``Unconditioned" (output of unconditional Music Transformer). We collected a total of 492 ratings, with each source involved in 328 distinct pair-wise comparisons.

\subsubsection{Melody and Performance Conditioning}
For melody and performance conditioning, we similarly presented participants with a 20s performance clip from the YouTube evaluation dataset and a 20s melody from a different piece in the evaluation dataset that we used as our conditioning signals. We then asked each participant to listen to two additional 20s performance clips and to use a Likert scale to rate which sounded most like the conditioning melody played in the style of the conditioning performance. The sources rated by the participants included ``Melody \& Performance" (output of the Melody-Performance Autoencoder), ``Melody only" (output of a model conditioned only on the melody signal), ``Performance only" (output of a model conditioned only on the performance signal), and ``Unconditioned" (output of an unconditional model). For this study, we collected a total of 714 ratings, with each source involved in 357 distinct pair-wise comparisons.

\Figref{fig:listening_tests} shows the number of comparisons in which each source was selected as being most similar in style to the conditioning signal. A Kruskal-Wallis H test of the ratings showed that there is at least one statistically significant difference between the models: $\chi^2(2) = 332.09,~p < 0.05~(7.72\mathrm{e}{-73})$ for melody conditioning and $\chi^2(2) = 277.74,~p < 0.05~(6.53\mathrm{e}{-60})$ for melody and performance conditioning. A post-hoc analysis using the Wilcoxon signed-rank test with Bonferroni correction showed that there were statistically significant differences between all pairs of the performance study with $p < 0.05/3$ and all pairs of the performance and melody study with $p < 0.05/6$ except between the ``Melody only" and ``Melody \& Performance" models ($p = 0.0894$).

These results demonstrate that the performance conditioning signal has a clear, robust effect on the generated output: in the 164 comparisons between ``Ground Truth" and ``Conditioned", participants responded that they had a preference for ``Conditioned" sample 58 times.

Although the results between ``Melody-only" and ``Melody \& Performance" are close, this study demonstrates that conditioning with both melody and performance outperforms conditioning on performance alone, and they are competitive with melody-only conditioning, despite the model having to deal with the complexity of incorporating both conditioning signals. In fact, we find quantitative evidence that human evaluation is more sensitive to perceptual melodic similarity, as the ``Performance-only" model performs worst -- a slight contrast to the results from the OA metric in Section 5.2.

Our qualitative findings from the audio examples and interpolations, coupled with the quantitative results from the OA similarity metric and the listening test which capture different aspects of the synthesized performance, support the finding that the Melody \& Performance autoencoder offers significant control over the generated samples. We provide several audio examples demonstrating the effectiveness of these conditioning signals in the online supplement at \exampleswebpage.

%% file: sections/related.tex
\section{Related Work}
\label{related}
\textbf{Measuring music similarity:} We note that quantifying music similarity is a difficult problem. We incorporate and extend upon the rich line of work for measuring music similiarity in symbolic music \citep{ghias1995query,berenzweig2004large,slaney2008learning,hung2019improving,yang2018evaluation} for our setting, in which we evaluate similarities between polyphonic piano performances as opposed to monophonic melodies.

\textbf{Sequential autoencoders:} Building on the wealth of autoencoding literature \citep{hinton2006reducing,salakhutdinov2009deep,vincent2010stacked}, our work bridges the gap between the traditional sequence-to-sequence framework \citep{sutskever2014sequence}, their recent advances with various attention mechanisms \citep{vaswani2017attention,shaw2018self,huang2018music}, and sequential autoencoders. Though \citep{wangt} propose a Transformer-based conditional VAE for story generation, the self-attention mechanism is shared between the encoder and decoder. Most similar to our work is that of \citep{kaiser2018discrete}, which uses a Transformer decoder and a discrete autoencoding function to map an input sequence into a discretized, compressed representation. We note that this approach is complementary to ours, where a similar idea of discretization may be applied to the output of our Transformer encoder. The MusicVAE \citep{roberts2018hierarchical} is a sequential VAE with a hierarchical recurrent decoder, which learns an interpretable latent code for musical sequences that can be used during generation time. This work builds upon \citep{bowman2015generating} that uses recurrence and an autoregressive decoder for text generation. Our Transformer autoencoder can be seen as a deterministic variant of the MusicVAE, with a complex self-attention mechanism based on relative positioning in both the encoder and decoder architectures to capture more expressive features of the data at both the local and global scale.

\textbf{Controllable generations using representation learning:} There is also considerable work on controllable generations, where we focus on the music domain. \citep{engel2017latent} proposes to constrain the latent space of unconditional generative models to sample with respect to some predefined attributes, whereas we explicitly define our conditioning signal in the data space and learn a global representation of its style during training. The Universal Music Translation network aims to translate music across various styles, but is not directly comparable to our approach as they work with raw audio waveforms \citep{mor2018universal}. Both \citep{meade2019exploring} and MuseNet \citep{payne2019} generate music based on user preferences, but adopt a slightly different approach: the models are specifically trained with labeled tokens (e.g., composer and instrumentation) as conditioning input, while our Transformer autoencoder's global style representation is learned in an unsupervised way. We emphasize Transformer autoencoder's advantage of learning unsupervised representations of style, as obtaining ground truth annotations for music data may be prohibitively challenging.

%% file: sections/conclusion.tex
\section{Conclusion}
\label{conclusion}
We proposed the Transformer autoencoder for conditional music generation, a sequential autoencoder model which utilizes an autoregressive Transformer encoder and decoder for improved modeling of musical sequences with long-term structure. We show that this model allows users to easily adapt the outputs of their generative model using even a single input performance. Through experiments on the MAESTRO and \dataset datasets, we demonstrate both quantitatively and qualitatively that our model generates samples that sound similar in style to a variety of conditioning signals relative to baselines. For future work, it would be interesting to explore other training procedures such as variational techniques or few-shot learning approaches \citep{finn2017model,reed2017few} to account for situations in which the input signals are from slightly different data distributions than the training set. We provide open-sourced implementations in Tensorflow \citep{abadi2016tensorflow} at \texttt{https://goo.gl/magenta/music-transformer-\\autoencoder-code}.

%% file: sections/acks.tex
\section*{\normalsize{Acknowledgements}}
We are thankful to Anna Huang, Hanoi Hantrakul, Aditya Grover, Rui Shu, and the Magenta team for insightful discussions. KC is supported by the NSF GRFP, QIF, and Stanford Graduate Fellowship.

%% file: sections/supplement.tex
\section*{Supplementary Material}
\renewcommand{\thesubsection}{\Alph{subsection}}
\label{appendix}

\subsection{Additional Details on Melody Representation}
For the melody representation (vocabulary), we followed \citep{waite2016} to encode the melody as a sequence of 92 unique tokens and quantized it to a 100ms grid. For the extraction procedure, we used the algorithm as outlined in the Magenta codebase (\texttt{https://github.com/tensorflow/magenta/\\blob/master/magenta/music/\\melody\_inference.py}), where we use a heuristic to extract the note with the highest in a given performance. This heuristic is based on the assumption that all melodies coincide with actual notes played in the polyphonic performance. Specifically, we construct a transition matrix of melody pitches and use the Viterbi algorithm to infer the most likely sequence of melody events within a given frame.

\subsection{NLL Evaluation for "Noisy" Model}
Below, we provide the note-wise test NLL on the MAESTRO and \dataset datasets with melody conditioning, where the conditioning performance is perturbed by the procedure outlined in Section~\ref{c_generation}.

\begin{table*}[ht]
\centering
\begin{tabular}{l|l|l}
\toprule
\textbf{Model variation} & MAESTRO & YouTube Dataset \\
\midrule
Noisy Melody TF autoencoder with relative attention, sum & 1.721 & 1.248\\   
Noisy Melody TF autoencoder with relative attention, concat & 1.719 & 1.249\\   
Noisy Melody TF autoencoder with relative attention, tile & 1.728 & 1.253\\   
\bottomrule
\end{tabular}
\caption{Note-wise test NLL on the MAESTRO and \dataset piano performance datasets with melody conditioning, with event-based representations of lengths $L = 2048$.}
\label{table:melody_nll_noisy}
\end{table*}

\subsection{Model Architecture and Hyperparameter Configurations}

We mostly use the default Transformer architecture as provided in the Tensor2Tensor framework, such as 8 self-attention heads as listed in the main text, and list the slight adjustments we made for each dataset below:

\subsubsection{MAESTRO}
For the MAESTRO dataset, we follow the hyperparameter setup of \citep{huang2018music}:
\begin{enumerate}
\item num hidden layers = 6
\item hidden units = 384
\item filter size = 1024
\item maximum sequence length = 2048
\item maximum relative distance = half the hidden size
\item dropout = 0.1
\end{enumerate}

\subsubsection{\dataset Dataset}
For the \dataset dataset, we modify the number of hidden layers to 8 and slightly increase the level of dropout.
\begin{enumerate}
\item num hidden layers = 8
\item hidden units = 384
\item filter size = 1024
\item maximum sequence length = 2048
\item maximum relative distance = half the hidden size
\item dropout = 0.15
\end{enumerate}

\subsection{Additional Relative Distance Interpolations}
\begin{figure}[ht]
\centering
\subfigure[Relative distance from interpolated sample to the original starting performance.]{\includegraphics[width=.45\textwidth]{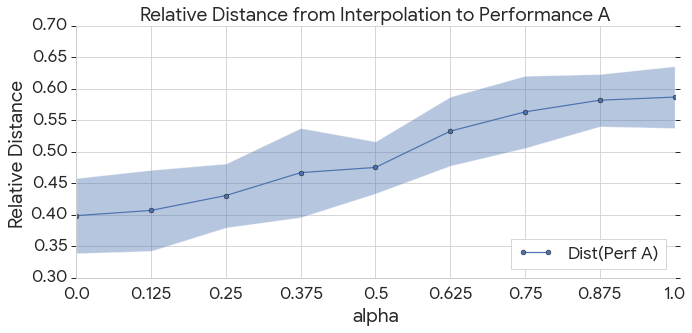}}
\subfigure[Relative distance from the interpolated sample to the original melody, which is kept fixed.]{\includegraphics[width=.45\textwidth]{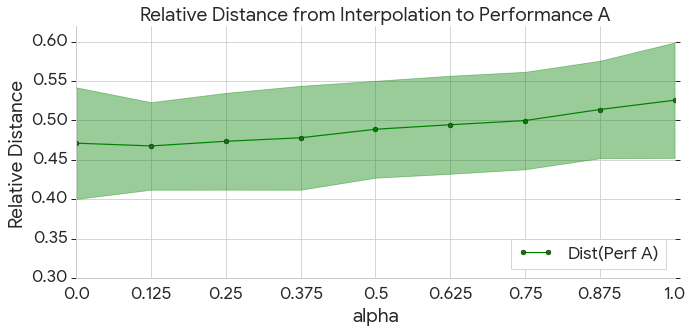}}
\caption{The distance to the original performance increases as the value of $\alpha$ increases in (a), as expected. In (b), we see that there is a very slight increase in the relative distance to the original melody during the interpolation procedure.}
\label{fig:interp_dist_mel_supp}
\end{figure}

In Figure~\ref{fig:interp_dist_mel_supp}, we show the interpolation relative distance results for the (a) performance and (b) melody \& performance Transformer autoencoders for the MAESTRO dataset.

We find consistent results in these interpolations as provided in the main text.

\subsection{Internal Dataset Performance Interpolations}
\begin{figure*}[ht]
\centering

\subfigure[Original starting performance]{\includegraphics[width=.49\textwidth]{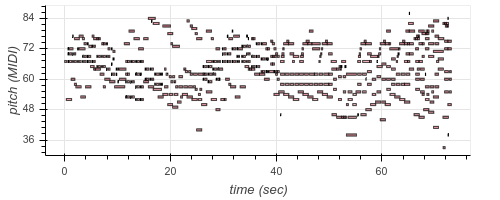}}
~
\subfigure[$\alpha=0.125$]{\includegraphics[width=.49\textwidth]{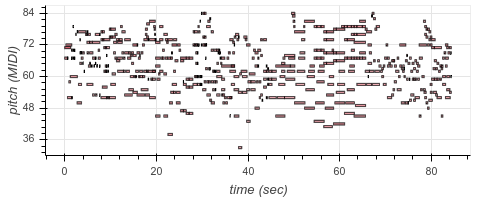}}
~
\subfigure[$\alpha=0.375$]{\includegraphics[width=.49\textwidth]{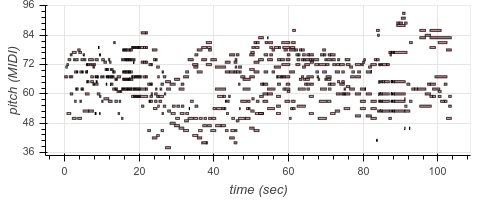}}
~
\subfigure[$\alpha=0.5$]{\includegraphics[width=.49\textwidth]{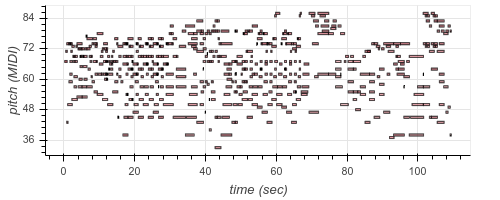}}
~
\subfigure[$\alpha=0.625$]{\includegraphics[width=.49\textwidth]{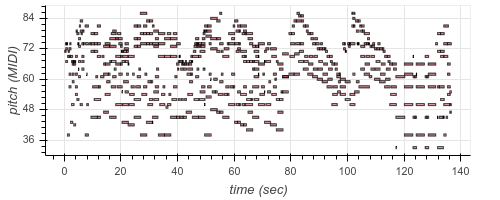}}
~
\subfigure[$\alpha=0.875$]{\includegraphics[width=.49\textwidth]{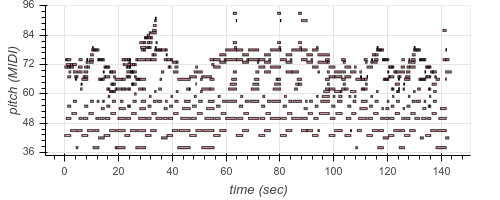}}
~
\subfigure[$\alpha=1.0$ (reconstruction)]{\includegraphics[width=.49\textwidth]{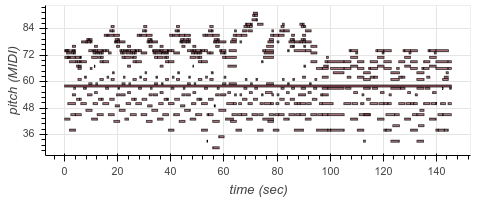}}
~
\subfigure[Original final performance]{\includegraphics[width=.49\textwidth]{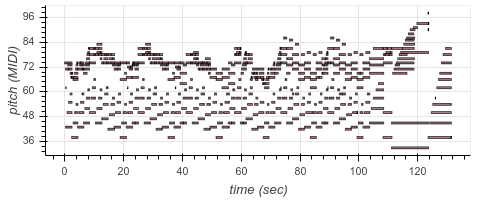}}
\caption{Interpolation of a starting performance (a) from the \dataset dataset to a final performance (h), with the coefficient $\alpha$ controlling the level of interpolation between the latent encodings between the two performances.}
\label{fig:interp_latent_perf}
\end{figure*}

\begin{figure*}[ht]

\centering
\subfigure[Original melody]{\includegraphics[width=.49\textwidth]{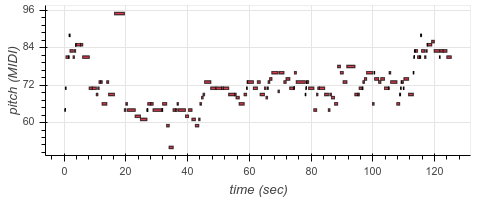}}
~
\subfigure[Original starting performance]{\includegraphics[width=.49\textwidth]{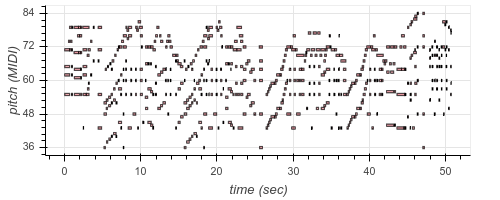}}
~
\subfigure[$\alpha=0$ (reconstruction)]{\includegraphics[width=.49\textwidth]{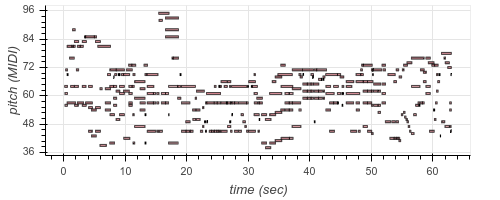}}
~
\subfigure[$\alpha=0.125$]{\includegraphics[width=.49\textwidth]{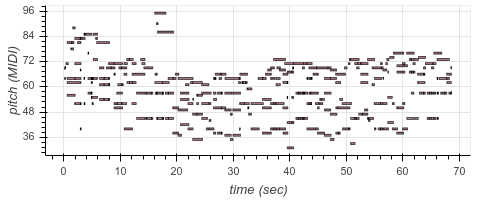}}
~
\subfigure[$\alpha=0.375$]{\includegraphics[width=.49\textwidth]{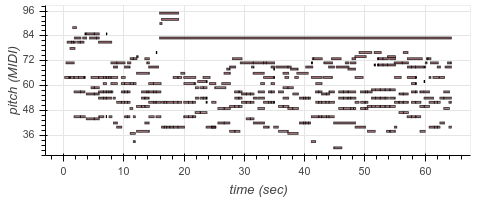}}
~
\subfigure[$\alpha=0.5$]{\includegraphics[width=.49\textwidth]{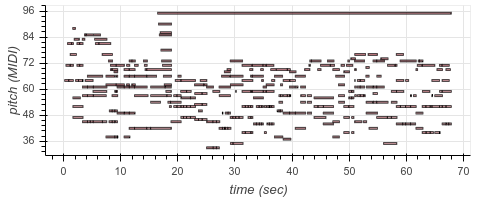}}
~
\subfigure[$\alpha=0.625$]{\includegraphics[width=.49\textwidth]{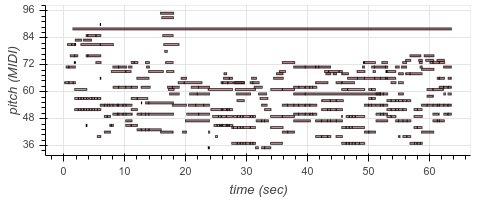}}
~
\subfigure[$\alpha=0.875$]{\includegraphics[width=.49\textwidth]{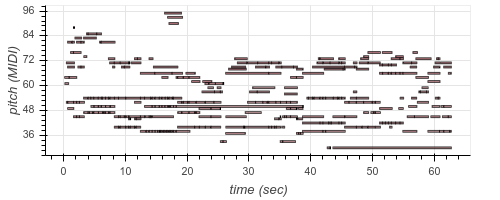}}
~
\subfigure[$\alpha=1.0$ (reconstruction)]{\includegraphics[width=.49\textwidth]{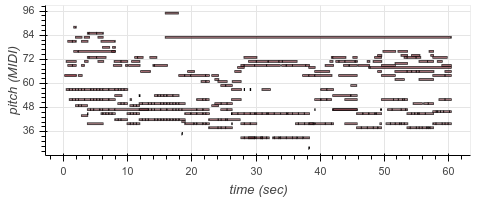}}
~
\subfigure[Original final performance]{\includegraphics[width=.49\textwidth]{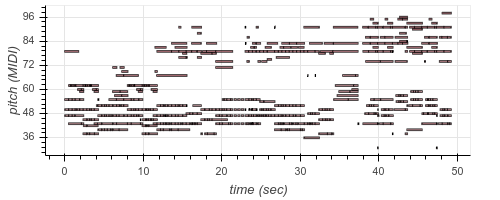}}

\caption{Interpolation of a starting performance (b) from the \dataset dataset to a final performance (j), with the coefficient $\alpha$ controlling the level of interpolation between the latent encodings between the two performances. The original conditioning melody (a) is kept fixed throughout the interpolation.}
\label{fig:interp_latent_mel}
\end{figure*}

In Figures~\ref{fig:interp_latent_perf} and~\ref{fig:interp_latent_mel}, we provide piano rolls demonstrating the effects of latent-space interpolation for the \dataset dataset, for both the (a) performance and (b) melody \& performance Transformer autoencoder respectively. For similar results in MAESTRO as well as additional listening samples, we refer the reader to the online supplement: \exampleswebpage.

\pagebreak

